\documentclass[10pt]{article}

\usepackage{color}
\usepackage{latexsym}
\usepackage{amsmath, amssymb, amsfonts, amsthm}
\usepackage{dsfont}
\usepackage{pdfsync}

\usepackage[a4paper]{geometry}
\usepackage{yfonts}

\newtheorem{Theorem}{Theorem}[part]
\newtheorem{Definition}{Definition}[part]
\newtheorem{Proposition}{Proposition}[part]
\newtheorem{Assumption}{Assumption}[part]
\newtheorem{Standing Assumption}{Standing Assumption}

\newtheorem{Corollary}{Corollary}[part]

\makeatletter \@addtoreset{equation}{section}

\@addtoreset{Definition}{section}

\@addtoreset{Theorem}{section}

\@addtoreset{Proposition}{section}

\@addtoreset{Assumption}{section}

\@addtoreset{Corollary}{section}

\@addtoreset{Lemma}{section}

\@addtoreset{Remark}{section}

\@addtoreset{Example}{section}

\makeatother \makeatletter

\def \F{\mathbb{F}}
\def \G{\mathbb{G}}

\def \P{\mathbb{P}}
\def \Q{\mathbb{Q}}
\def \R{\mathbb{R}}

\def \we{\widetilde}

\def \Bc{{\cal B}}

\def \Fc{{\cal F}}
\def \Gc{{\cal G}}

\def \Mc{{\cal M}}

\def \Qc{{\cal Q}}

\def \Tc{{\cal T}}

\def \proof{{\noindent \bf Proof}\quad}
\def \ep{\hbox{ }\hfill$\Box$}
\def \reff#1{{\rm(\ref{#1})}}
\def \[{[\,\!\![}
\def \]{]\,\!\!]}
\def \beqn{\begin{eqnarray}}
\def \eeqn{\end{eqnarray}}
\def \beq{\begin{equation}}
\def \eeq{\end{equation}}
\def \beqn*{\begin{eqnarray*}}
\def \eeqn*{\end{eqnarray*}}
\def \ri#1{\mbox{ri}(#1)}

\def \x{\times}

\def \cone#1{{\rm cone}(#1)}
\def \conv{\mbox{\rm conv}}

\def \={\;=\;}

\def \implies{\Rightarrow}

\def \And{\;\mbox{ and }\;}

\def \pourtout{\mbox{ for all } }
\def \st{\mbox{ s.t. }}
\def \Pas{\mathbb{P}-\mbox{a.s.}}

\def \.{\;.}

\def \we{\widetilde}
\def \wt{\widehat}

\def \eps{\varepsilon}

\def \vs#1{\vspace{#1mm}}
\def \1{{\bf 1}}
\def \0{{\bf 0}}
\def \ind#1{\mathds{1}_{\brace{#1}}}

\def \Esp#1{\mathbb{E}\left[#1\right]}

\def \Pro#1{\mathbb{P}\left[{#1}\right]}

\def \NMA{\mbox{\rm{\bf NMA}}}

\def \ti{{t_i}}
\def \tip{ {t_{i+1}} }
\def \tim{ {t_{i-1}}}

\def \eqref#1{\reff{#1}}

\def \p{+}

\def \cadlag{c\`adl\`ag }

\def \ladlag{l\`adl\`ag }

\def \Xf{\mathfrak{X}}

\def \Kf{\mathfrak{K}}
\def \brace#1{\left\{ #1 \right\}}
\def \implies{\Longrightarrow}

\def \equival{\Longleftrightarrow}

\title{A note on super-hedging for investor-producers}

\author{ 
 Adrien Nguyen Huu \thanks{This research is part of the Chair {\it Finance and Sustainable Development} sponsored by EDF and CACIB. } 
 						\\\small CEREMADE - Univ. Paris-Dauphine
						\\\small and Finance for Energy Market Research Centre
                        \\\small Paris, France
                        \\\small nguyen@ceremade.dauphine.fr}

\begin{document}

\maketitle

\begin{abstract}
We study the situation of an agent who can trade on 
a financial market and can also transform
some assets into others by means of a production system,
in order to price and hedge derivatives on produced goods.
This framework is motivated by the case of an electricity producer 
who wants to hedge a position on the electricity spot price and 
can trade commodities which are inputs for his system.
This extends the essential results of \cite{bouchard2011nma} 
to continuous time markets.
We introduce the generic concept of \emph{conditional sure profit}  
along the idea of the \emph{no sure profit} condition of R\'{a}sonyi \cite{rasonyi2009aut}.
The condition allows one to provide
a closedness property for the set of
super-hedgeable claims in a very general financial setting.
Using standard separation arguments, 
we then deduce a dual characterization of the latter
and provide an application to power futures pricing.
\end{abstract}

\noindent {\bf Key words~:} 
arbitrage pricing theory,
markets with proportional transaction costs, 
non-linear returns, 
super replication theorem, 
electricity markets, 
energy derivatives.

\section{Introduction}
\label{sec:introduction}

The recent deregulation of electricity markets in many countries 
has opened a new range of applications for 
financial techniques in order to hedge energy risks.
However,
the non-storability of electricity forbids
any trading strategy based on the spot price and
the standard mathematical toolbox cannot be exploited 
to hedge and price derivative products upon this asset.
The challenge must still be taken up for 
electricity producers who are endowed with such claims.
It also concerns
financial agents possessing a power plant,
as an asset
for diversification purposes.
These economical agents can produce power out of
a storable commodity,
and sell it to benefit from 
electricity prices variation.
Hence, they perform a sort of financial strategy,
as studied in \cite{aid2009asr}
where the production is
a structural function of electricity demand.
Here, 
our goal is to study the general situation of an agent 
who can trade financial assets and inputs for his production system, 
and who can transform a position into an other by the mean of
a production system he controls.

As in \cite{bouchard2011nma},
the reasoning is the following.
In the framework of purely financial portfolios, 
Arbitrage Pricing Theory ensures by an economical assumption, 
the no-arbitrage condition, 
a closedness property for the set of 
attainable terminal wealth for self financing portfolios.
This key property has direct applications, 
such as a dual formulation, 
which provides an equivalent martingale measure
for pricing purposes.
In our particular framework,
if the financial market runs as usual, 
production is not bound up with any particular economical condition :
it is an idiosyncratic action of the agent.
We thus propose in this note a general constraint
upon the production possibilities of the agent
in order to apply arbitrage pricing techniques.
In practice,
the additional condition is calibrated to market data
and the producer's activity.
In theory,
this condition implies
the closedness property of the set of attainable terminal positions,
as it is sought in the purely financial case.
This property allows to display
many financial techniques,
such as risk measures or portfolio optimization.
The purpose of this note is to demonstrate and apply 
the undermentioned super-replication theorem for the investor-producer.
If we denote $\Xf_0^R (T)$ the set of 
possible portfolio outcomes at time $T$
that the investor-producer can reach starting from $0$ at time $0$,
and $\Mc$ the set of \textit{pricing measures}
for the financial market model,
we thus show afterwards the following result:
\begin{Theorem}\label{th:SR}
Let $H$ be a contingent claim.
Then
$$H \in \Xf^R_{0}(T)\ \equival \ \Esp{Z'_T H}\le \alpha_{0}^R(Z), 
\ \forall Z\in \Mc$$
where 
$
\alpha_{0}^R(Z):=\sup\brace{\Esp{Z'_T V_T}~:~V_T \in \Xf^R_{0}(T)}
$ 
is the support function of $Z\in \Mc$ on $\Xf^R_{0}(T)$.
\end{Theorem}

The usual interpretation is that a contingent claim is replicable
with a strategy starting from nothing at time $0$ if and only if the expectation
with respect to a pricing measure $Z\in \Mc$ always verifies a given bounding condition. 
The paper is thus structured around that theorem as follows.
In Section \ref{sec: model},
we introduce properly 
the entities $\Xf_0^R (T)$, $\Mc$ and $H$.
In Section \ref{sec: CSP}, 
we propose the economical condition 
under which Theorem \ref{th:SR} holds.
In Section \ref{sec:application}, 
we give an application to Theorem \ref{th:SR}.
Section \ref{sec:proof} is dedicated to 
the proof of Theorem \ref{th:SR}.

This problem has actually been explored 
in a discrete time framework 
for markets with proportional transaction costs 
in \cite{bouchard2011nma}.
In the latter, 
the authors propose to extend
the no-arbitrage of second kind condition of R\'{a}sonyi \cite{rasonyi2009aut} 
to portfolios augmented by a linear production system.
A condition for general production functions, 
the \emph{no marginal arbitrage for high production regime} condition, 
has then been introduced using the extended condition above 
in order to allow marginal arbitrages for reasonable levels of production.
In the present note,
we push forward 
this study by proposing an alternative condition 
which has a close economical interpretation:
the \textit{conditional sure profit} condition.
Contrary to the \textit{no marginal arbitrage condition} 
of \cite{bouchard2011nma}, 
it deals directly with general production possibilities and
avoids to introduce a linear production system.
This is the contribution of Section \ref{sec: CSP}.
We also focus on investors-producers with 
specific means of production:
production possibilities are in discrete time 
as in \cite{bouchard2011nma} but 
we additionally assume concavity and boundedness of the production function.
In counterpart, 
our framework encompasses continuous time financial market models
with and without transaction costs.
This is the contribution of Section \ref{sec: model}.
The contribution of Section \ref{sec:application} is
to apply Theorem \ref{th:SR}
in order to put a price on a power futures contract
for an electricity producer endowed with a simple mean of production. 

\vs2

{\bf General notations}: 
throughout this note, 
$x\in \R^d$ will be viewed as a column vector 
with entries $x^i$, $i\le d$.
The transpose of a vector $x$ will be denoted $x'$, 
so that $x'y$ stands for the scalar product.
As usual, 
$\R^d_+$ and $\R^d_-$ stand for the positive and negative orthants of $\R^d$ respectively, i.e., $[0, +\infty)^d$ and $(-\infty, 0]^d$.
For a given probability space $(\Omega, \Gc, \P)$ 
and a $\Gc$-measurable random set $E$, 
$L^0(E, \Gc)$ will denote the set of 
$\Gc$-measurable random variables 
taking values in $E$ $\P$-almost surely, 
$L^1(E, \Gc)$ the set of 
$\P$-integrable random variables 
takings values in $E$ $\P$-almost surely and 
$L^\infty(E, \Gc)$ the set of 
essentially bounded random variables taking values in $E$.
The notation $\conv(E)$ will denote 
the closed convex hull of $E$,
and $\cone {E}$ the closed convex cone generated by $\conv(E)$.
All the inclusions or inequalities are to be understood 
in the almost sure sense unless otherwise specified.

\section{The framework}
\label{sec: model}

We first introduce the financial possibilities of the agent.
We consider an abstract setting
mainly inspired by \cite{denis2010cps}, 
which allows to deal with a very large class of market models.
To illustrate our framework, 
we provide two examples in Section \ref{sec: examples}.
We then introduce production possibilities for the investor.

\vs2

\textbf{Preamble.}
Let $(\Omega, \Fc, \F=(\Fc_t)_{t\in [0,T]}, \P)$ be a continuous-time 
filtered stochastic basis on a finite time interval $[0,T]$ satisfying the usual conditions.
We assume without loss of generality that $\Fc_0$ is trivial and $\Fc_{T^-} = \Fc_T$.
For any $0\le t\le T$, 
let $\Tc$ denote the family of 
stopping times taking values in $[0,T]$ $\P$-almost surely.
From now on, 
we consider a pair of set-valued $\F$-adapted process 
$\wt K$ and  $\wt K^*$ such that 
$\wt K_t (\omega)$ is a proper convex closed cone of $\R^d$
including $\R^d_+$,
$dt\otimes d\P$-a.e.
The process $\wt K^*$ is defined by 
\begin{equation}\label{eq:dualcone}
\wt K^*_t (\omega) :=\brace{y\in \R^d_+~:~ xy\ge0 , \; \forall x\in \wt K_t(\omega)} \; .
\end{equation}
Since $\wt K_t(\omega)$ is proper, its dual 
$\wt K^*_t (\omega)\ne \brace{0}$ $dt\otimes d\P$-a.s.
In the literature on markets with transaction costs, 
$\wt K_t$ usually stands for the solvency region at time $t$, 
and $-\wt K_t$ for the set of possible trades at time $t$, 
see \cite{kabanov2010book} and the reference therein.
In practice, 
$\wt K$ and $\wt K^*$ are given by the market model we consider, 
see the examples of Sections \ref{sec: examples} and \ref{sec:application}.
We use here the process $\wt K$ to introduce a partial order on $\R^d$ 
at any stopping time in $\Tc$.
\begin{Definition}\label{def: partial order}
Let $\tau \in\Tc$. 
For $(\xi, \kappa)\in L^0(\R^{2d}, \Fc_\tau)$,
$\xi \succeq_\tau -\kappa$ if and only if 
$\xi + \kappa \in L^0(\wt K_\tau, \Fc_\tau)$.
\end{Definition}

\begin{Definition}\label{def: contingent claim}
A contingent claim is a random variable $H\in L^0(\R^d, \Fc_T)$ such that 
$H\succeq_T -\kappa$ for some $\kappa \in \R_+^d$. 
\end{Definition}

\subsection{The set of financial positions}

We consider a financial market on $[0,T]$ with $d$ assets.
The market also includes the prices of commodity
entering in the production process, 
e.g., fuel or raw materials.
The agent we consider has the possibility to trade on this market
by starting a portfolio strategy at any time $\rho\in \Tc$.
The financial possibilities of the agent
are then represented by a family of sets of 
wealth processes denoted $(\Xf_{\rho}^{0})_{\rho \in \Tc}$.
The superscript $0$ stands for \emph{no production}, 
or \emph{pure financial}.

\begin{Definition}\label{def: portfeuille}
For any $\rho \in \Tc$,
the set $\Xf_{\rho}^{0}$ is a set of 
$\F$-adapted $d$-dimensional processes $\xi$ 
defined on $[0,T]$ such that
$\xi_{t}=0 \ \Pas$ for all $t \in [0,\rho)$.
We denote by
$\Xf^0_{\rho}(T):=\brace{\xi_T~:~\xi \in \Xf^0_{\rho}}$ 
the corresponding set of attainable financial positions at time $T$.
\end{Definition}

We do not give more details on what a financial strategy is.
In all the considered examples,
it will denote a self financing portfolio value
as commonly defined in Arbitrage Pricing Theory.
The multidimensional setting is justified by 
models of financial portfolios in markets 
with proportional transaction costs, 
see \cite{kabanov2010book}.
In that case,
portfolio are expressed in physical units of assets.
Just note that we implicitly assume that
the initial wealth of the agent does not influence
his financial possibilities,
so that a portfolio generically starts with 
a null wealth in our setting.


\begin{Assumption}\label{ass: portefeuille}
For any $\rho \in \Tc$, 
the set  $\Xf_{\rho}^{0}(T)$ has the following properties:
\begin{enumerate}
\item[(i)] \emph{Convexity}: $\Xf_{\rho}^{0}(T)$ is a convex subset of $L^0(\R^d, \Fc_T)$ containing $0$.
\item[(ii)]\label{eq:liquidation} \emph{Liquidation possibilities}: $
\Xf^0_{\rho}(T) - L^\infty (\wt K_s, \Fc_s)\subseteq \Xf^0_{\rho}(T), \quad \forall s\in [\rho, T] \ \Pas
$ 
\item[(iii)]\label{eq:concatenation} \emph{Concatenation}: $
\Xf^0_{\rho}(T)=\brace{\xi_\sigma +\zeta_T~:~ (\xi,\zeta)\in \Xf^0_{\rho}\x \Xf^0_{\sigma}, \text{ for any }\sigma \in \Tc \text{ s.t. }\sigma \ge \rho}\; .$
\end{enumerate}
\end{Assumption}

The convexity property holds in most of market models, 
see \cite{kabanov2010book}.
Assumption \ref{eq:liquidation}.(ii) means that 
whatever the financial position of the agent is, 
it is always possible for him to throw away 
a non-negative quantity of assets at any time, 
or to do an arbitrarily large transfer of assets 
allowed by the cone $-\wt K_s$.
This last possibility is made for models
of markets with convex transaction costs.
Finally, 
the concatenation property also holds 
in most of market models and often 
reveals their Markovian behaviour.
Note that
Assumption \reff{eq:concatenation} (i) and (iii) imply that
$
\Xf^0_\rho (T)\subset \Xf^0_\tau (T)
$ 
for any $(\rho, \tau)\in \Tc^2$ such that $\rho\ge \tau$.

\subsection{Absence of arbitrage in the financial market}

As for any investor on a financial market,
we assume that our investor-producer cannot find an arbitrage opportunity.
We elaborate below this condition
by relying on the core result of Arbitrage Pricing Theory,
which resides in the following fact,
see the introduction of \cite{denis2010cps}.
Formally, 
when the financial market prices 
are represented by a process $S$, 
the no-arbitrage property for the market holds 
if and only if there exists a stochastic deflator, 
i.e., a strictly positive martingale $\Gamma$ such that 
the process $Z:=\Gamma S$ is a martingale.
The process $Z$ can then be seen as 
the shadow price or fair price of assets.
We assume that such a process $Z$ exists
by introducing the following.

\begin{Definition}\label{def:M}
Let $\Mc$ be the set of $\F$-adapted martingales 
$Z$ on $[0,T]$ taking values in $\wt K^*$, 
with strictly positive components,
such that
\begin{equation}\label{eq: support fini}
\sup \brace{\Esp{Z'_T \xi_T} \; : \; \xi \in \Xf^0_{0} 
\And 
\exists \kappa \in \R^d_+ \st
\forall \tau\in \Tc,
\xi_\tau \succeq_\tau -\kappa
}<+\infty\; .
\end{equation}
\end{Definition}
In condition \reff{eq: support fini},
we apply the pricing measure $Z$ for the subset of $\Xf_0^0 (T)$
comprising financial wealth processes with a finite credit line $\kappa$.
We need this basic concept of 
admissibility for portfolio processes to define $\Mc$
properly.
We will extend admissibility of wealth processes
in the next section.
Definition \ref{def:M} needs more comment.
If the set $\Xf^0_{0}(T)$ is a cone,
the left hand of \reff{eq: support fini} 
is null for any $Z\in \Mc$,
according to Assumption \ref{ass: portefeuille} (i).
In the general non conical case, 
see Section \ref{ex:2}, 
the support function in equation \reff{eq: support fini} might be positive,
justifying the more general condition.
If it is equal to $0$ then,
for any $Z\in \Mc$ and
any $\xi\in\Xf_{0}^{0}$ with a finite credit line,
according to Assumption \ref{ass: portefeuille} (iii),
$Z'\xi$ is a supermartingale.
We then meet the common no arbitrage condition,
see especially Section \ref{ex:1} below.
We thus express absence of arbitrage on 
the financial market by the following assumption.
\begin{Assumption}
$\Mc \ne \emptyset$.
\end{Assumption}
Note that 
defining $\Mc$ as above is tailor-made for separation arguments, 
see the proof of Theorem \ref{th:SR}. 

\subsection{Admissible portfolios and closedness property}

If $d=1$, 
a financial position $\xi_t$ is naturally solvable if 
$\xi_t \ge 0 $ $\P-$a.s.
In the general setting with $d\ge 1$, 
we use the partial order on $\R^d$ induced by the process $\wt K$. 
Defining solvency allows to define admissibility
which is central in continuous time:
the closedness property concerns the subset of $\Xf_0^0(T)$
constituted of admissible portfolios,
see \cite{delbaen1994agv, campi2006asr, denis2010cps, denis2011tft}
and the various definitions provided therein.
From a financial point of view, 
it imposes realistic constraints on portfolios
and avoids doubling strategies.
Here, we use a definition close to the one proposed in \cite{campi2006asr}.
\begin{Definition}\label{def:admissibility}
For some constant vector $\kappa\in \R^d_+$, a portfolio $\xi\in \Xf^0_{0}$ is said to be 
$\kappa$-\emph{admissible} if $Z'_\tau \xi_\tau \ge -Z'_\tau \kappa$ 
for all $\tau\in \Tc$ and all $Z\in \Mc$, and $\xi_T \succeq_T - \kappa$.
\end{Definition}
Given $\Mc\ne \emptyset$, 
the concept of admissibility allows to consider 
a wider class of terminal wealth than 
those considered in equation \reff{eq: support fini}. 
According to Definition \ref{def:M},
a wealth process $\xi$ is $\kappa$-admissible 
in the sense of Definition \ref{def:admissibility}
if $\xi$ verifies $\xi_\tau\succeq_\tau -\kappa$
for all $\tau\in \Tc$ and some $\kappa\in \R^d_+$.
The reciprocal is not always true, and is the object of the so-called {\bf B} assumption
investigated in \cite{denis2010cps}.
We can finally define the set of admissible elements of $\Xf_t^0$:
\begin{Definition}\label{def: admissible portefeuille}
We define
$
\Xf_{t,adm}^0 := \brace{\xi \in \Xf_t^0, \ \xi \text{ is }\kappa\text{-admissible for some }\kappa\in \R^d_+}
$,
and $\Xf_{t,adm}^0 (T):=\brace{\xi_T~:~\xi \in \Xf_{t, adm}^0}$.
\end{Definition}
The closedness property will be assigned to the sets $\Xf_{t,adm}^{0}(T)$,
and is conveyed under 
the following technical and standing assumption:

\begin{Assumption}\label{ass:convergence}
For $t\in [0,T]$, 
let $(\xi^n)_{n\ge 1} \subset \Xf^0_{t, adm}$ be 
a sequence of admissible portfolios such that 
$\xi^n_T\succeq_T -\kappa$ for some $\kappa\in \R^d_+$ and all $n\ge 1$.
Then there exists 
a sequence $ (\zeta^n)_{n\ge 1}\subset \Xf^0_{t, adm}$ constructed as 
a convex combination 
(with strictly positive weights) 
of $(\xi^n)_{n\ge 1} $, 
i.e., $\zeta^n\in \conv(\xi^k)_{k\ge n}$,  
such that $\zeta_T^n$ converges a.s. to 
$\zeta^{\infty}_T\in \Xf^0_{t}(T)$ with $n$.
\end{Assumption}

The above assumption calls for the notion of Fatou-convergence.
Recall that a sequence of random variables is Fatou-convergent if 
it is bounded by below 
and almost surely convergent.
According to Assumption \ref{ass: portefeuille} (i),
$\Xf_{t,adm}^0(T)$ is a convex set, 
which ensures that the new sequence lies in the set.
In Arbitrage Pricing Theory,
the Fatou-closedness of $\Xf^0_0(T)$ often relies on 
a convergence lemma.
Schachermayer \cite{schachermayer1992ahs} introduced 
a version of Komlos Lemma that is fundamental in \cite{delbaen1994agv},
while Campi and Schachermayer \cite{campi2006asr} 
proposed another version for markets with 
proportional transaction costs.
Assumption \ref{ass:convergence} expresses a synthesis of this result, 
see Sections \ref{ex:1} and \ref{sec:application} for applications.

\subsection{Illustration of the framework by examples of financial markets}
\label{sec: examples}

We illustrate here the theoretical framework. 
We treat two examples, 
based on \cite{delbaen1994agv, delbaen1995teo} 
and \cite{pennanen2009hop, kabanov2003otc} respectively.
In section \ref{sec:application}, 
we also apply our results to a continuous time market 
with \cadlag price processes and proportional transaction costs, 
as studied in \cite{campi2006asr}.

\subsubsection{A multidimensional frictionless market in continuous time}\label{ex:1}

Consider a filtered stochastic basis 
$(\Omega, \Fc, \F, \P)$ on $[0,T]$,
satisfying the usual assumptions.
Let $S$ be a locally bounded 
$(0,\infty)^d$-valued $\F$-adapted \cadlag semimartingale,
representing the price process of $d$ risky assets.
We suppose the existence of 
a non risky asset which is taken constant on $[0,T]$
without loss of generality.
Let $\Theta$ be the set of $\F$-predictable $S$-integrable processes and 
$\Pi$ the set of $\F$-predictable increasing processes on $[0,T]$. 
We define 
$$
\Xf^0_{\rho}:=\brace{\xi=(\xi^1, 0, \ldots, 0) \; :\; 
\xi^1_s = \int_\rho^s \vartheta_u .dS_u - (\ell_s - \ell_{\rho^-}) ~:~
(\vartheta, \ell)\in \Theta\x \Pi, \; s\in[\rho,T]}\; , \quad \forall \rho\in \Tc \; .
$$
Observe that the set
$\Xf^0_{\rho}(T)$ is a convex cone
of $\R\x \brace{0}^{d-1}$ containing $0$.
The set $\Theta$ defines the financial strategies.
The set $\Pi$ represents 
possible liquidation or consumption in the portfolio.
The introduction of the latter 
ensures Assumption \reff{eq:liquidation} (ii),
but does not infer on 
the mathematical treatment of \cite{delbaen1994agv} 
where $\Pi$ is not considered.
The set $\Xf^0_{0}(T)$ also verifies 
Assumption \reff{eq:liquidation} (i) and (iii).

\vs2

In this context,
Delbaen and Schachermayer introduced 
the No Free Lunch with Vanishing Risk condition (NFLVR)
and proved that it is equivalent to
$$
\Qc := \brace{ \Q \sim \P \text{ such that } S \text{ is a }\Q-\text{local martingale}}\ne \emptyset \quad \text{(Theorem 1.1 in \cite{delbaen1994agv}) } \.
$$
To relate the NFLVR condition to Definition \ref{def:M}, 
we define $\Mc$ as the set of 
$\P$-equivalent local martingale measure processes 
$\left. \frac{d\Q}{d\P}\right|_{\Fc_.}$
for $\Q\in \Qc$.
If $S$ is a locally bounded martingale, 
elements of $\Xf^0_0$ are local martingales.
We now apply Definition \ref{def:admissibility} of admissibility. 
We take without ambiguity 
$\wt K= \wt K^*=\R_+^d$.
As a consequence,
a portfolio $\xi\in\Xf_{0,adm}^0$ is $\kappa$-admissible 
only if $\xi^1_t\ge -\kappa$ for all $t\in [0,T]$,
and we retrieve 
the definition of admissibility of \cite{delbaen1994agv}.
Therefore, 
any admissible portfolio is 
a true supermartingale under $\Q\in \Qc$.

\vs2

By Theorem 4.2 in \cite{delbaen1994agv},
NFLVR implies that 
$\Xf^0_{0,adm}(T)$ is Fatou-closed.
The proof uses the following convergence property:
for any $1$-admissible sequence $\xi^n\in \Xf^0_{0}$, 
it is possible to find 
$\zeta^n \in \conv(\xi^k)_{k\ge n}$ such that 
$\zeta^n$ converges in the semimartingale topology 
(Lemmata 4.10 and 4.11  in \cite{delbaen1994agv}). 
Hence, $\zeta^n_T$ Fatou-converges in $\Xf^0_{0}(T)$.
This can be easily extended to $\Xf^0_{\tau}(T)$ 
for any $\tau \in \Tc$ and 
for any bound of admissibility.
In this case, 
Definition \ref{def:admissibility} and 
the martingale property of $\xi^1$ 
imply uniform admissibility in the sense of \cite{delbaen1994agv}.
Assumption \ref{ass:convergence} then holds in this context.

\subsubsection{A physical market with convex transaction costs in discrete time}\label{ex:2}

Let $(t_i)_{0\le i\le N}\subset [0,T]$ be 
an increasing sequence of deterministic times with $t_N=T$.
Let us consider 
the discrete filtration $\G:=(\Fc_\ti)_{0\le i \le N}$.
Here, 
the market is modelled by 
a $\G$-adapted sequence $C=(C_\ti)_{0\le i\le N}$ of 
closed-valued mappings $C_\ti :\Omega \mapsto \R^d$ with 
$\R^d_-\subset C_\ti (\omega)$ and $C_\ti(\omega)$ convex 
for every $0\le i\le N$ and $\omega\in \Omega$.
We define 
the recession cones 
$C^\infty_t (\omega) = \bigcap_{\alpha>0}\alpha C_t (\omega)$ and 
their dual cones 
$C^{\infty,* }_t(\omega)=\brace{y\in \R^d ~:~ xy\ge 0, \ \forall x\in C^\infty_t (\omega)}$, 
see also \cite{pennanen2009hop} for a freestanding definition

\vs2

This setting has been introduced in \cite{pennanen2009hop} 
to model markets with convex transaction costs, 
such as currency markets with illiquidity costs, 
in discrete time.
Every financial position is labelled in
 physical units of the $d$ assets, 
and the sets $C_\ti$ denote the possible 
self financing changes of position at time $\ti$, 
so that
$$
\Xf^0_{t_i}(T)
:=\brace{\sum_{k=i}^N \xi_{t_k} ~:~ \xi_{t_k} \in L^0(C_{t_k}, \Fc_{t_k}), 
\ \forall i\le k\le N}  \pourtout 0\le i\le N\.
$$
In this context, 
Assumption \ref{ass: portefeuille} trivially holds.
If $C_\ti(\omega)$ is a cone in $\R^d$ 
for all $0\le i\le N$ and $\omega\in \Omega$, i.e., $C = C^\infty$,
we retrieve a market with proportional transaction costs 
as described in \cite{kabanov2003otc}.
In the latter, 
Kabanov and al. show that 
the Fundamental Theorem of Asset Pricing 
can be expressed with respect to 
the \emph{robust no-arbitrage} property, 
see \cite{kabanov2003otc} for a definition.
This condition is equivalent to 
the existence of a martingale process $Z$ such that 
$Z_\ti \in L^\infty( \ri{ C^{\infty, *}_\ti}, \Fc_\ti )$, 
where $\ri{ C^{\infty, *}_\ti}$ denotes 
the relative interior of $ C^{\infty, *}_{\ti}$.
The super replication theorem, 
see Lemma 3.3.2 in \cite{kabanov2010book}, 
allows $\Mc$ given by 
Definition \ref{def:M} 
to be characterized by such elements $Z$.
In that case, 
the reader can see that $C^\infty$ replaces 
our conventional cone process $\wt K$.

\vs2

As mentioned in \cite{pennanen2009hop}, 
the case of general convex transaction costs 
leads to two possible definitions of arbitrage.
One of them is based on the recession cone.
Following the terminology of \cite{pennanen2009hop}, 
the market represented by $C$ satisfies 
the \emph{robust no-scalable arbitrage} property 
if $C^\infty$ satisfies the robust no-arbitrage property.
This definition implies that arbitrages might exist, 
but they are limited for elements of $\Xf^0_0(T)$ 
and even not possible for the recession cone.
Pennanen and Penner \cite{pennanen2009hop} proved that 
the set $\Xf^0_{0}(T)$ is closed in probability under this condition.
Hence, it is Fatou-closed.
The convergence result used in this context is 
a different argument than 
the one of Assumption \ref{ass:convergence}.
However, the latter can be applied, 
see \cite{bouchard2011nma} in which 
Assumption \ref{ass:convergence} has been applied in 
a very similar context.
The notion of admissibility can also be avoided 
in the discrete time case.

%

\subsection{Addition of production possibilities}

The previous introduction of 
a financial market comes from 
the possibility to interpret 
the available assets on the market
as raw material or saleable goods for a producer.
Therefore, 
we model the production as 
a function transforming a consumption of the $d$ assets 
in a new wealth in $\R^d$.
Other observations from 
the situation of an electricity provider 
lead to our upcoming setting.
On a deregulated electricity market, 
power is provided with respect to 
an hourly time grid.
Production control can thus be fairly approximated by 
a discrete time framework.
We also introduce a delay in the control, 
as a physical constraint in the production process.
See \cite{kallrath2009book} for 
a monograph illustrating these concerns.

\begin{Definition}\label{def: production}
Let $(t_i)_{0\le i\le N}\subset [0,T]$ be 
a deterministic collection of strictly increasing times.
We then define a \emph{production regime} as an element
$\beta$ in $\Bc$, where
$$
\Bc:=\brace{(\beta_\ti)_{0\le i <N} ~:~ 
\beta_\ti \in L^0(\R^d_+, \Fc_\ti), \  0\le i <N} \; .
$$
A \emph{production function} is then 
a collection of maps $R:=(R_\ti)_{0< i \le N}$ such that
for $0<i\le N$, 
$R_\ti$ is a $\Fc_\ti$-measurable map from $\R^d_+$ to $\R^d$,
in the sense that 
$R_{\ti}(\beta_\tim)\in L^0(\R^d, \Fc_\ti)$ 
for $\beta_\tim\in L^0(\R^d_+, \Fc_\tim)$.
\end{Definition}
Without loss of generality,
it is also possible to consider 
an increasing sequence of stopping times 
in $\Tc$ instead of the $(t_i)_{0\le i\le N}$.
The set $\Bc$ can also be defined via 
sequences $(\beta_\ti)_{0\le i<N}$ such that 
$\beta_\ti$ takes values in a convex closed subset of $\R^d_+$.
The proofs in section \ref{sec:proof} would be identical 
and we refrain from doing this.
Notice also that it has no mathematical cost to consider 
separate times of injection and times of production, i.e., 
a non-decreasing sequence 
$
\brace{t_0, s_0, t_1, s_1,\ldots, t_{N}, s_{N}}\subset [0,T]
$
with $t_i<s_i$, $(t_i)_{0\le i< N}$ and $(s_i)_{0< i\le N}$ allowing to define $\Bc$ and $R$ respectively.
As invoked in the introduction,
we add fundamental assumptions on the production function.
\begin{Assumption}\label{ass:R}
The production function has the three following properties:
\begin{itemize}
	\item[(i)] \emph{Concavity}: for all $0<i\le N$, for all $(\beta^1, \beta^2)\in L^0(\R^{2d}_+, \Fc_\tim )$ and $\lambda \in L^0([0,1],\Fc_\tim)$, 
    $$
     R_\ti(\lambda \beta^1\p (1-\lambda)\beta^2) - \lambda R_{\ti}(\beta^1) - (1-\lambda)R_\ti(\beta^2) \in \R^d_+ \ \Pas
    $$
	\item[(ii)] \emph{Boundedness}: there exists a constant $\Kf \in \R^d_+$ such that for all $0<i\le N$, 
	$$\Kf - |R_\ti(\beta) - \beta| \in \R^d_+ \; \Pas \;,\; \pourtout \beta\in\R^d_+ \; .$$
	\item[(iii)] \emph{Continuity}: For any $0<i\le N$, we have that
	$
	\lim\limits_{\beta^n \to \beta^0} R_\ti(\beta^n) = R_\ti (\beta^0) \; .
	$
\end{itemize}
\end{Assumption}

These assumptions are fundamental for the continuous time setting.
Assumption \ref{ass:R} (i) keeps 
the convexity property for the set $\Xf_0^R(T)$, 
see Proposition \ref{prop:convexity} in the proofs section.
Assumption \ref{ass:R} (ii) does not only ensure 
the admissibility of 
investment-production portfolios when we add production.
From the economical point of view,
it affirms that 
the net production income is bounded, 
which forbids infinite profits.
It thus provides a realistic framework 
for physical production systems.
Finally,
Assumption \ref{ass:R}.(iii) is a technical assumption
in order to use Assumption \ref{ass:convergence}.
It is only needed to ensures upper semicontinuity 
on the boundary of $\R^d_+$,
since continuity comes from (i) inside the domain.
See corollary 4.3 in \cite{bouchard2011nma}, 
where convexity is not needed and
upper semicontinuity is sufficient.

Notice that concavity and 
the upper bound $\Kf$ 
for the production incomes are given 
with respect to $\R^d_+$ and not $\wt K$.
This is a useful artefact in the proofs, 
but also a meaningful expression of a physical bound of production,
which has nothing to do with a financial model.
With Assumption \ref{ass:R},
it is possible to fairly approximate 
a generation asset, 
see Section \ref{sec:application}.

\vs2

\begin{Definition}\label{def: production portefeuille}
The set of investment-production wealth processes starting at time $t$ is given by
$$
\Xf^R_{t}:=\brace{V \; : \; V_s:=\xi_s+ \sum_{i=1}^N R_\ti (\beta_\tim\ind{\tim \ge t})\ind{\ti\le s} - \beta_\tim \ind{t\le \tim \le s}, \;  (\xi, \beta)\in \Xf^0_{t, adm}\x\Bc}  \.
$$
The set of terminal possible outcomes for the investor-producer is given by $\Xf_t^R (T):=\brace{V_T~:~V\in \Xf^R_{t}}$.
\end{Definition}

The agent manages his production system as follows.
Assume that he starts an investment-production strategy at time $t$.
On one hand, 
he performs a financial strategy given by $\xi \in \Xf^0_{t, adm}$.
On the other hand,
he can decide to put 
a quantity of assets $\beta_\tim$ at time $\ti$ 
into the production system if $\tim \ge t$.
The latter returns a position $R_\ti (\beta_\tim)$ 
labelled in assets at time $\ti$.
At this time, 
the agent also decides the regime of production $\beta_\ti$ 
for the next step of time, 
and so on until time reaches $t_N$.


\vs2

The generalization to 
continuous time controls raises 
mathematical difficulties.
When coming to a continuous time control, 
we have to make a distinction between 
the continuous and the discontinuous part of the control, 
i.e., 
between a regime of production as a rate 
and an instantaneous consumption of 
assets put in the production system.
This natural distinction has already been observed 
for liquidity matters in financial markets, 
see \cite{cetin2006poi}.
This implies a separate treatment of consumption in the function $R$.
With a continuous control and as in \cite{cetin2006poi}
the production becomes a linear function of that control,
which is very restrictive and similar to
the polyhedral cone setting of
markets with proportional transaction costs.
With a discontinuous control,
non linearity can appear but
we face two difficulties.
If the number of discontinuities is bounded,
it is easy to see that 
the set of controls is not convex.
On the contrary,
if it is not bounded,
the set is not closed.
This problem typically appears in
impulse control problems and is not easy to overcome,
see Chapter 7 in \cite{oksendal2005applied}.
We ought to focus on that difficulty in another paper.

\section{The conditional sure profit condition}
\label{sec: CSP}

In the situation of our agent, 
even if we accept no arbitrage on 
the financial market,
there is no economical justification for 
the interdiction of 
profits coming from the production.
This is the reason why the concept of 
\emph{no marginal arbitrage for high production regime} 
has been introduced in \cite{bouchard2011nma} ($\NMA$ for short).
The $\NMA$ condition expresses 
the possibility to make sure profits 
coming from the production possibilities, 
but that marginally tend to zero 
if the production regime $\beta$ is pushed toward infinity.
This condition relied on 
an affine bound for the production function, 
introducing then an auxiliary 
linear production function for which 
sure profits are forbidden.
We propose another parametric condition based on
the idea of possibly making solvable profits 
for a small regime of production.
It is stronger than $\NMA$ under Assumption \ref{ass:R}, 
see Remark 2.5 in \cite{bouchard2011nma}, but 
we express directly 
the new condition with the production function.

\begin{Definition}\label{def:na}
We say that there are only 
\emph{conditional sure profits} for the production function $R$,  
{\bf CSP}(R) holds for short,
if there exists $C>0$ such that 
for all $0\le k< N$ and 
for all $(\xi,\beta )\in \Xf^0_{t_k, adm}\x \Bc$ 
we have:
$$
\xi_T + \sum_{i=k}^{N-1} R_\tip(\beta_\ti) - \beta_\ti  \succeq_T \sum_{i=k}^{N-1} R_\tip(0) \ \Pas \quad \Longrightarrow \quad  \|\beta_\ti\|\le  C \text{ for }k\le i< N \; .
$$
\end{Definition}

The condition {\bf CSP}(R) thus reads as follows.
If the agent starts an investment-production strategy at 
an intermediary date $t\in (t_{k-1}, t_k]$ 
for some $k$ (whatever his initial position is at $t$), 
then he can start his production at index $k$.
We then assess that he can do better than 
the strategy $(0,0)\in \Xf^0_0 \x \Bc$ 
only if the regime of production is bounded.
{\bf CSP}(R) comes from the following observation:
coming back to the usual case of a financial market, 
a possible interpretation of a no-arbitrage condition is that 
there is no strategy which is better than the null strategy  $\P$-a.s.
We thus transpose this interpretation to production with a slight modification.
Here, doing nothing implies that the agent is subject to 
possible fixed costs expressed by $R(0)$.
Since we do not specify portfolios by an initial holding,
we can focus on portfolios starting 
at any time before $T$ with any initial position.

\vs2

The terminology {\bf CSP}(R) refers to 
the \emph{no sure profit} property 
introduced by Rasonyi \cite{rasonyi2009aut}
(which became the \emph{no sure gain in liquidation value} 
condition in the final version),
since it is formulated in a very similar way and expresses 
the interdiction for sure profit if some condition is not fulfilled.
The {\bf CSP}(R) property is indeed very flexible.
It is possible to change 
the condition ``$\|\beta_\ti\|\le C$ for $k\le i\le N$'' by any restriction of the form:
$$
\text{``There exists a value } c_i \in (0,+\infty) 
\st 
\|\beta_\ti\| \ne c_i  \pourtout 0\le i <N \; \text{''} .
$$
This can convey the condition that 
the regime of production shall be null or greater than 
a threshold to allow profits, 
or observe a more precise condition on its components
as long as it also constrains the norm of $\beta$.
Posing {\bf CSP}(R) implies that
the closedness property on the financial market alone
transmits to the market with production possibilities.
Theorem \ref{th:SR} given in introduction then follows
as a corollary to the following proposition.

\begin{Proposition}\label{prop:FC}
The set $\Xf^R_{0}(T)$ is Fatou-closed under {\bf CSP}(R).
\end{Proposition}

\section{Application to the pricing of a power future contract}
\label{sec:application}

We illustrate Theorem \ref{th:SR} by 
an application to an electricity producer 
endowed with a generation system converting a raw material, 
e.g. fuel, into electricity and who has 
the possibility to trade that asset on a market.
We address here the question of a possible price of 
a term contract a producer can propose on power
when he takes into account his the generation asset.
We assume that the financial market is submitted 
to proportional transaction costs.
For this reason, 
we place ourselves in the financial framework 
developed by Campi and Schachermayer \cite{campi2006asr}.

\subsection{The financial market}

We consider a financial market on $[0,T]$ 
composed of two assets, 
cash and fuel, 
which are indexed by $1$ and $2$ respectively.
The market is represented by a so-called bid-ask process $\pi$, 
see \cite{campi2006asr} for a general definition.

\begin{Assumption}\label{ass:bid-ask}
The process $\pi=(\pi^{12}_t, \pi^{21}_t)_{t}$ is a $(0,+\infty)^2$-valued
$\F$-adapted \cadlag process verifying efficient frictions, i.e.,
$$
\pi^{12}_t \x \pi^{21}_t>1 \pourtout t\in [0,T] \; \Pas 
$$
\end{Assumption}
Here $\pi^{12}_t$ denotes at time $t$ 
the quantity of cash necessary to obtain 
and $(\pi^{21}_t)^{-1}$ denotes 
the quantity of cash that can be obtained 
by selling one unit of fuel.
The efficient frictions assumption conveys
the presence of positive transaction costs.
The process $\pi$ generates a set-valued random process which defines the solvency region:
$$
\wt K_t(\omega) := \cone{ e^1, e^2, \pi^{12}_t(\omega) e^1 - e^2, \pi^{21}_t(\omega) e^2 - e^1} \quad \forall (t,\omega)\in [0,T]\x\Omega \;.
$$
Here $(e^1, e^2)$ is the canonical base of $\R^2$.
The process $\wt K$ is $\F$-adapted and closed convex cone-valued.
It provides the partial order on $\R^2$ of Definition \ref{def: partial order}.
\begin{Assumption}\label{ass:campi portfolio}
Every $\xi \in\Xf_0^0$ is a \ladlag $\R^2$-valued
$\F$-predictable process with finite variation verifying,
for every $(\sigma, \tau)\in  \Tc^2_{[0,T]}$ with $\sigma< \tau$,
$$
(\xi_\tau -\xi_\sigma)(\omega) \in \overline{\conv} \left(  \bigcup_{\sigma(\omega) \le u \le \tau(\omega)}-\wt K_u (\omega) \right) \;,
$$
the bar denoting the closure in $\R^d$.

\end{Assumption}
Assumption \ref{ass:campi portfolio} implies Assumption \ref{ass: portefeuille}.
Admissible portfolios are defined via Definitions \ref{def:admissibility}
and \ref{def: partial order}.

\begin{Corollary}\label{cor:M}
Every $Z\in \Mc$ is a $\R^2_+$-valued martingale verifying
$(\pi^{21}_t)^{-1} \le Z^1_t / Z^2_t \le \pi^{12}_t$ $\Pas$
and:
\begin{itemize}
\item for all $\sigma\in \Tc$, 
$(\pi^{21}_\sigma)^{-1} < Z^1_\sigma / Z^2_\sigma < \pi^{12}_{\sigma}$;
\item for all predictable $\sigma\in \Tc$, 
 $(\pi^{21}_{\sigma^-})^{-1} < Z^1_{\sigma^-} / Z^2_{\sigma^-} < \pi^{12}_{\sigma^-}$.
\end{itemize}
\end{Corollary}

\proof
The market model is conical, 
so that 
$
\alpha^0_{0} (Z) := \sup\brace{\Esp{Z'_T V_T}~:~ V\in \Xf^0_{0,adm}}=0$,
for all $Z\in \Mc$.
The fact that Definition \ref{def:M} corresponds to these elements $Z$
follows from the construction of $\wt K$ and is 
a part of the proof of Theorem 4.1 in \cite{campi2006asr}.
\ep

\vs2

Under the assumption that $\Mc \ne \emptyset$,
$Z\xi$ is a supermartingale for all $Z\in \Mc$ and
admissible $\xi\in \Xf_0^0$, 
see Lemma 2.8 in \cite{campi2006asr}.
Finally, 
Assumption \ref{ass:convergence} is given by Proposition 3.4 in \cite{campi2006asr}.
For a comprehensive introduction of all these objects, 
we refer to \cite{campi2006asr}.

\subsection{The generation asset}

We suppose that the agent possesses a thermal plant 
allowing to produce electricity  out of fuel 
on a fixed period of time.
The electricity spot price is determined per hour, 
so that we define the calendar of production as $(t_i)_{0\le i\le N}\subset [0,T]$,
where $N$ represents the number of generation actions for each hour of the fixed period.
At time $\ti$, 
the agent puts a quantity $\beta_\ti=(\beta^1_\ti, \beta^2_\ti)$ of assets in the plant.
The production system transforms at time $\tip$ the quantity $\beta^2_\ti$ of fuel, 
given a fixed heat rate $q_{i+1}\in \R_+ $, 
into a quantity $q_{i+1} \beta^2_\ti$ of electricity (in MWh).
The producer has a limited capacity of injection of fuel given by 
a threshold $\Delta_{i+1}\in L^\infty(\R_+, \Fc_\tip)$.
This implies that any additional quantity over $\Delta_{i+1}$ 
of fuel injected in the process will be redirected to storage facilities, 
i.e., as fuel in the portfolio.
The electricity is immediately sold on the market via 
the hourly spot price.
On most of electricity markets, 
the spot price is legally bounded.
It can also happen to be negative.
It is thus given by 
$
P_{i+1}\in L^\infty (\R, \Fc_\tip)
$.
For a given time $\tip$, 
the agent is subject to a fixed cost $\gamma_{i+1}$ in cash.
The agent also faces a cost in fuel in order to maintain the plant activity.
This is given by a supposedly non-positive increasing concave function $c_{i+1}$ on $[0,\Delta_{i+1}]$ such that 
 $c'_{i+1} (\Delta_{i+1})\ge 1$, 
where $c'_{i+1}$ represents the left derivative. 
Altogether, we propose the following.

\begin{Assumption}\label{ass:production function}
The production function is given by
$
R_\tip (\beta_\ti) = (R^1_\tip (\beta_\ti), R^2_\tip(\beta_\ti))
$
for $0 \le i <N$, where
$$
\left\{\begin{array}{ccl}
R^1_\tip ((\beta^1_\ti,\beta_\ti^2)) 
&=& P_{i+1} q_{i+1} \min(\beta^2_\ti, \Delta_{i+1} ) - \gamma_{i+1} + \beta^1_\ti\\
R^2_\tip((\beta^1_\ti,\beta^2_\ti))
&=& c_{i+1} (\min (\beta^2_\ti, \Delta_{i+1})) + 
\max(\beta^2_\ti - \Delta_{i+1}, 0)
\end{array}
\right. \; .
$$
\end{Assumption}

We can constraint $\beta^1_\ti$ to be null at every time $\ti$ without any loss of generality.

\begin{Corollary}\label{cor:bonnes hyp}
Assumption \ref{ass:R} holds under Assumption \ref{ass:production function}.
\end{Corollary}

\proof
For each $i$, $R_{\tip}$ verifies Assumption \ref{ass:R} (ii):
$$
|R^1_\tip((\beta^1_\ti,\beta_\ti^2)) - \beta^1_\ti| \le  |P_{i+1} q_{i+1}\Delta_{i+1}| + |\gamma_{i+1}| \in L^\infty(\R, \Fc_\tip) 
$$
and 
$$
|R^2_\tip((\beta^1_\ti,\beta_\ti^2)) - \beta^2_\ti|\le \max (|c_{i+1}(0)|,| c_{i+1}(\Delta_{i+1}) - \Delta_{i+1} | )  \in L^\infty(\R_+, \Fc_\tip) \; .
$$
Notice that since $c_{i+1}$ is concave 
with $c'_{i+1}(\Delta_{i+1})\ge 1$, 
the function $R_{\tip}^2$ is clearly concave.
The function $R$ is then concave 
in each component with respect to the usual order,
so that Assumption \ref{ass:R} (i) holds 
with the partial order induced by $\wt K$.
It is also continuous, 
so that Assumption \ref{ass:R} (iii) holds.
\ep

\subsection{Super replication price of a power futures contract}

We now fix a condition provided by the agent 
in order to apply Definition \ref{def:na}.
For example suppose that the agent knows at time $\ti$ that
by producing under a typical regime $C$ 
(a given threshold of fuel to put in his system)
and selling the production at the market price, 
he can refund the quantity of fuel needed to produce.
It is a conceivable phenomenon on the electricity spot market.
Since the electricity spot price is actually an increasing function 
of the total amount of electricity produced by the participants, 
the agent can sell a small quantity of electricity at high price
if the total production is high.
He can then partially or totally recover his fixed cost and even make sure profit.
The constant $C$ can depend on external factors of the model, 
such as the level of aggregated demand of electricity.

\begin{Assumption}\label{ass:arbitrage}
We assume that there exists $C>0$ such that
\begin{equation}\label{eq:arbitrage}
R^1_\tip(\beta^2_\ti) + \gamma_{i+1}\ge  (\pi^{12}_\tip)^{-1} (R^2_\tip (\beta^2_\ti) - \beta^2_\ti - c_{i+1}(0))  \ \Pas \implies \beta^2_\ti \le C
\end{equation}
\end{Assumption}

Here,
an immediate transfer $\xi_{\tip}$ of 
quantity $R^1_\tip(\beta^2_\ti)$ of asset 1
brought in asset $2$ gives 
$\xi_\tip + R_{\tip}(\beta_\ti) \succeq R_\tip (0)$, 
and {\bf CSP}(R) condition holds under Assumption \ref{ass:arbitrage}.
The latter thus implies that 
the set $\Xf^R_{0,adm}(T)$ is Fatou-closed, 
so that we can apply Theorem \ref{th:SR}.

\vs2

Now we consider the following contingent claim.
We denote by $F(x)$ the price of 
a power futures contract with physical delivery.
Buying this contract at time $0$ provides 
a fixed power $x$ (in MW) for 
$N$ consecutive hours of a fixed period.
Here, 
the $N$ hours correspond to the $(\ti)_{1\le i\le N}$.
Theorem \ref{th:SR} can be immediately applied to obtain the price at which
the investor-producer can sell the contract.

\begin{Corollary}\label{cor:price}
The price is given by 
$
F(x)= \sup_{Z\in \Mc} \left(\frac{1}{Z^1_0} \Esp{\sum_{i=1}^{N}  Z^1_\ti P_i x}- \alpha^R_{0}(Z)  \right)
$
where 
$$
\alpha^R_{0} (Z) = \sup_{\beta\in \Bc}\Esp{ \sum_{i=1}^N Z^1_\ti \left(P_{i}q_i \min(\beta^2_\tim, \Delta_i) - \gamma_i\right) 
+ Z^2_\ti  \left( c_i (\min(\beta^2_\tim, \Delta_i)) -  \min(\beta^2_\tim, \Delta_i) \right)} \; .
$$
\end{Corollary}

The theorem then ensures the existence of 
a wealth process, 
involving a financial strategy 
starting with wealth $F(x)$ and production activities,
such that
his terminal position is solvent $\Pas$.


\section{Proofs}
\label{sec:proof}

\subsection{Proof of Proposition \ref{prop:FC}}

We define a collection of sets
$$
\we \Xf^k_{t}:=\brace{V~:~V_s := \xi_s + \sum_{i=1}^{k} R_{t_{N+1-i}}(\beta_{t_{N-i}})\ind{t_{N+1-i}\le s}  - \beta_{t_{N-i}}\ind{t_{N-i}\le s} , \  (\xi,\beta)\in \Xf^0_{t,adm}\x \Bc}
$$
and
$
\we \Xf^k_{t}(T) :=\brace{V_T~:~ V\in \we \Xf^k_t } 
$
for $t\in [0,T]$ and $0\le k \le N$, 
with the convention that 
$$
\sum_{i=1}^0 R_{t_{N+1-i}}(\beta_{t_{N-i}}) - \beta_{t_{N-i}}=0 \; .
$$
Note thus that $\we \Xf^0_{t}(T)$ corresponds precisely 
to the set $\Xf_{t,adm}^0(T)$.
We are conducted by the following guideline.
According to Assumption \ref{ass:convergence}, 
$\we \Xf^0_{t_N}(T)$ is Fatou closed.
We then proceed by induction in two steps: 
we first show that $\we \Xf^k_{t_{N-(k+1)}}(T)$ is closed 
if $\we \Xf^k_{t_{N-k}}(T)$ is closed.
Then we prove that $\we \Xf^{k+1}_{t_{N-(k+1)}}(T)$ is closed if $\we \Xf^k_{t_{N-(k+1)}}(T)$ is closed.

\begin{Proposition}\label{prop:convexity}
For all $0\le k\le N $, the set $\we \Xf^k_{t_{N-k}}(T)$ is convex.
\end{Proposition}

\proof
This is a consequence of Assumption \ref{ass:R} (i).
Indeed take $(\xi^1, \beta^1)$ and $(\xi^2, \beta^2)$ in $\Xf^0_{t_{N-k}, adm}\x \Bc$ 
and $\lambda\in [0,1]$.
Take $(\kappa^1, \kappa^2) \in\R^{2d}_+$ the respective bounds of admissibility for $\xi^1$ and $\xi^2$.
Note that $\lambda \xi^1 + (1-\lambda)\xi^2$ is clearly 
$(\lambda \kappa^1 + (1-\lambda)\kappa^2)$-admissible 
since $\wt K$ is a cone-valued process.
By Assumption \ref{ass:R} (i), 
there exists $(\ell_{t_{N+1-i}})_{1\le i\le k}$ 
with $\ell_{t_{N+1-i}}\in  L^0(\R_-^d, \Fc_{t_{N+1-i}})$ such that 
$$
R_{t_{N+1-i}}(\lambda \beta^1_{t_{N-i}} + (1-\lambda)\beta^2_{t_{N-i}})+\ell_{t_{N+1-i}} = 
\lambda R_{t_{N+1-i}}(\beta^1_{t_{N-i}}) + (1-\lambda) R_{t_{N+1-i}}(\beta^2_{t_{N-i}})\;, \quad 1\le i\le k\; .
$$
Notice that $\R^d_- \subset \wt K_{t}$ for any $t\in [0,T]$, 
so that $\ell_{t_{N+1-i}}\in  L^0(-\wt K_{t_{N+1-i}}, \Fc_{t_{N+1-i}})$.
We will use this fact throughout the proof.
Notice also that, 
according to Assumption \ref{ass:R} (ii), 
each $\ell_{t_{N+1-i}}$ is bounded by below by $2\Kf$ for $1\le i\le k$,
where $\Kf$ is the bound of net production incomes.
By relation \reff{eq:liquidation} and the above fact,
$\lambda \xi_T^1+(1-\lambda)\xi_T^2 +\sum_{i=1}^k \ell_{t_{N+1-i}} \in \Xf^0_{t_{N-k}, adm}(T)$.
Assembling the parts gives the proposition.
\ep

\begin{Proposition}\label{prop:FC1}
If $\we \Xf^k_{t_{N-k}}(T)$ is Fatou-closed, then the same holds for $\we \Xf^k_{t_{N-(k+1)}}(T)$.
\end{Proposition}

\proof
Let $(V_T^n)_{n\ge 1}\subset \we \Xf^k_{t_{N-(k+1)}}(T)$ be a sequence such that 
$V^n_T$ Fatou-converges to some $V^0_T$.
Let $(\xi^n)_{n\ge 1}\subset \Xf^0_{t_{N-(k+1)}, adm}$ and $(\beta^n_{t_{N-i}})_{1\le i \le k, n\ge 1}$ 
with $(\beta^n_{t_{N-i}})_{n\ge 1}\subset L^0(\R^d_+, \Fc_{t_{N-i}})$ for $1\le i \le k$, 
and $\kappa\in \R^d_+$, 
such that 
$$
V^n_T = \xi^n_T + \sum_{i=1}^k R_{t_{N+1-i}} (\beta^n_{t_{N-i}}) - \beta^n_{t_{N-i}}\succeq_T - \kappa  \quad \forall n\ge 1 \; .
$$

According to Assumption \ref{ass:R} (ii), 
and since $\R^d_+ \subset \wt K_T$, 
we have that for any $n\ge 1$,
$$
-k\Kf \preceq_T \sum_{i=1}^k R_{t_{N+1-i}} (\beta^n_{t_{N-i}}) - \beta^n_{t_{N-i}}=: \wt V^n_T \in  \we \Xf^k_{t_{N-k}}(T)\; .
$$
Due to Assumption \ref{ass:R} (ii) also, 
we have that
$
\xi^n_T  \succeq_T -(\kappa + k \Kf)
$
for all $n\ge1$.
According to Assumption \ref{ass:convergence},
 we can then find a sequence of convex combinations $\we \xi^n$ of $\xi^n$, 
$\we{\xi}^n \in \conv (\xi^m)_{m\ge n}$, such that 
$\we{\xi}^n_T$ Fatou-converges to some $\we \xi^0_T\in \Xf^0_{t_{N-(k+1)}, adm}(T)$.
The convergence of $\we \xi^n_T$ implies, 
by using the same convex weights, 
that there exists a sequence $(\we V^n_T)_{n\ge 1}$ of 
convex combinations of $\wt V^m_T$, $m\ge n$, converging $\Pas$ to some $\we V^0_T$.
By Proposition \ref{prop:convexity} above, 
the sequence $(\we V^n_T)_{n\ge 1}$ lies in $\we \Xf^k_{t_{N-k}}(T)$.
Recall that it is also bounded by below.
Since $\we \Xf^k_{t_{N-k}}(T)$ is Fatou-closed, 
$\we V^0_T\in \we \Xf^k_{t_{N-k}}(T)$ and moreover, 
$\we V^0_T$ is of the form $\sum_{i=1}^k R_{t_{N+1-i}} (\beta^0_{t_{N-i}}) - \beta^0_{t_{N-i}} + \ell^0_{t_{N+1-i}}$
for some $\beta^0\in \Bc$ and $(\ell^0_{t_{N+1-i}})_{1\le i \le k}$ 
with $\ell^0_{t_{N+1-i}} \in L^\infty(-\wt K_{t_{N+1-i}}, \Fc_{t_{N+1-i}})$
for $1\le i \le k$.
This is due to Assumption \ref{ass:R} (i)-(ii).
If we let $(\lambda_m)_{m\ge n}$ be the above convex weights, 
we can always write for $1\le i \le k$ and $n\ge 1$
$$
\sum_{m\ge n} \lambda_m \left( R_{t_{N+1-i}}(\beta^m_{t_{N-i}}) -\beta^m_{t_{N-i}}\right)  = R_{t_{N+1-i}}(\sum_{m\ge n} \lambda_m \beta^m_{t_{N-i}}) - \sum_{m\ge n} \lambda_m\beta^m_{t_{N-i}} + \ell^n_{t_{N+1-i}}  \; .
$$
The sets $L^0(-\wt K_{t_{N+1-i}}, \Fc_{t_{N+1-i}})$ and $L^0(\R^d_+, \Fc_{t_{N-i}})$
are closed convex cones for $1\le i\le k$, 
so that $\ell^n_{t_{N+1-i}}$ and $\sum_{m\ge n} \lambda_m\beta^m_{t_{N-i}}$
and their possible limits stay in those sets respectively.
From the boundedness condition of Assumption \ref{ass:R} (ii), 
the vectors $\ell^n_{t_{N+1-i}}$ are uniformly 
bounded by below by $2\Kf$ for any $1\le i\le k$ and $n\ge1$, 
and so are $\ell^0_{t_{N+1-i}}$ for $1\le i\le k$.
According to \reff{eq:liquidation}, 
$\we \xi^n_T + \sum_{i=1}^k \ell^0_{t_{N+1-i}}\in \Xf^0_{t_{N-(k+1)}, adm}(T)$.
We then have that $\we \xi^n_T + \we V^N_T$ converges to 
$\we \xi^0_T + \we V^0_T = V^0_T\in\we \Xf^k_{t_{N-(k+1)}}(T)$.
\ep

\begin{Proposition}\label{prop:FC2}
If $\we \Xf^{k}_{t_{N-(k+1)}}(T)$ is Fatou-closed, then the same holds for $\we \Xf^{k+1}_{t_{N-(k+1)}}(T)$.
\end{Proposition}

\proof
Let $(V_T^n)_{n\ge 1}\subset \we \Xf^{k+1}_{t_{N-(k+1)}}(T)$ such that 
there exists $\kappa\in \R^d_+$ verifying $V_T^n\succeq_T -\kappa$ for $n\ge 1$, 
and $V_T^n$ converges $\Pas$ toward $V_T\in L^0(\R^d, \Fc_T)$ when $n$ goes to infinity.
We let $(\bar V^n_T, \bar \beta^n)_{n\ge 1}\subset \we \Xf^{k}_{t_{N-(k+1)}}(T)\x L^0(\R^d_+, \Fc_{t_{N-(k+1)}})$ 
be such that $V_T^n = \bar V_T^n + R_{t_{N-k}}(\bar\beta^n) - \bar\beta^n$.
Define $\eta^n=|\bar\beta^n|$ and the $\Fc_{t_{N-(k+1)}}$-measurable set $E:=\brace{\limsup_{n\rightarrow \infty}\eta^n <+\infty}$.
We consider two cases.

{\bf 1.}
First assume that $E=\Omega$.
Then $(\bar\beta^n)_{n\ge 1}$ is $\Pas$ uniformly bounded.
According to Lemma 2 in \cite{kabanov2001atn}, 
we can find a $\Fc_{t_{N-(k+1)}}$-measurable random subsequence of $(\bar\beta^n)_{n\ge 1}$, 
still indexed by $n$ for sake of clarity, 
which converges $\Pas$ to some $\bar\beta^0\in L^\infty(\R^d_+, \Fc_{t_{N-(k+1)}})$.
By Assumption \ref{ass:R} (iii), 
$R_{t_{N-k}}(\bar{\beta}^n)$ converges to $R_{t_{N-k}}(\bar\beta^0)$, 
Recall that $\bar V_T^n \succeq -\kappa - \Kf$ for $n\ge 1$.
Since it is $\P$-almost surely convergent to $V_T-R_{t_{N-k}}(\bar\beta^0) + \bar\beta^0 =:\bar V^0_T$ 
and that $\we \Xf^{k}_{t_{N-(k+1)}}(T)$ is Fatou-closed, 
the limit $\bar V^0_T$ lies in that set.
This implies that
 $V_T\in \we \Xf^{k+1}_{t_{N-(k+1)}}(T)$.

{\bf 2.}
Assume now that $\Pro{E^c}>0$.
Since $E^c$ is $\Fc_{t_{N-(k+1)}}$-measurable,
 we argue conditionally to that set 
and suppose without loss of generality that $E^c = \Omega$.
We then know that there exists a $\Fc_{t_{N-(k+1)}}$-measurable subsequence 
of $(\eta^n)_{n\ge 1}$ converging $\P$-almost surely to infinity 
with $n$ by an argument similar to the one of Lemma 2 in \cite{kabanov2001atn}.
We overwrite $n$ by the index of this subsequence.
We write $V_T^n$ as follows:
\begin{equation}\label{eq:forme utile}
V_T^n = \xi^n_T + R_{t_{N-k}}(\beta_{t_{N-(k+1)}}^n)- \beta_{t_{N-(k+1)}}^n + \sum_{i=1}^k R_{t_{N+1-i}}(\beta^n_{t_{N-i}}) - \beta^n_{{t_{N-i}}} \; ,
\end{equation}
with 
$
(\xi^n)_{n\ge 1} \subset \Xf^0_{t_{N-(k+1)}, adm}
$ 
and 
$
(\beta^n_{t_{N-i}})_{1\le i \le k+1, n\ge 1}
$ 
with 
$
(\beta^n_{t_{N-i}})_{n\ge 1}\subset L^0(\R^d_+, \Fc_{t_{N-i}})
$ 
for $1\le i \le k+1$, 
and with the natural convention that for all $n\ge 1$, $\beta_{t_{N-(k+1)}}^n = \bar\beta^n$.
We then define 
\begin{equation}\label{eq:normalization}
(\we V_T^n, \we \xi_T^n, \we \beta_{t_{N-(k+1)}}^n, \ldots, \we \beta_{t_N}^n) := \frac{2\|C\|}{1+\eta^n} (V_T^{n}, \xi_T^{n}, \beta_{t_{N-(k+1)}}^{n}, \ldots, \beta_{t_N}^n) \.
\end{equation}
Now that $(\we \beta_{t_{N-(k+1)}}^n)_{n\ge 1}$ is a bounded sequence, 
we can extract a random subsequence, still indexed by $n$, 
such that $(\we \beta_{t_{N-(k+1)}}^n)_{n\ge 1}$ converges $\Pas$ to 
$
\beta_{t_{N-(k+1)}}^0\in L^0(\R^d_+, \Fc_{t_{N-(k+1)}}) \;.
$
Notice for later that $\|\we \beta_{t_{N-(k+1)}}^n \|$ converges to $\| \beta_{t_{N-(k+1)}}^0\| = 2 \|C\|$.
It is clear that Assumption \ref{ass:R} (i) allows to write
\begin{equation}\label{eq:limite de R}
\frac{2\|C\|}{1+\eta^n} \left(R_{t_{N+1-i}} (\beta^n_{t_{N-i}}) - \beta^n_{t_{N-i}}\right) 
= 
R_{t_{N+1-i}} (\we \beta^n_{t_{N-i}}) - \we \beta^n_{t_{N-i}} - \left(1- \frac{2\|C\|}{1+\eta^n}\right)R_{t_{N+1-i}} (0) + \ell_{t_{N+1-i}}^n \;,
\end{equation}
with $(\ell^n_{t_{N+1-i}})_{n\ge 1} \subset L^\infty(-\wt K_{t_{N+1-i}}, \Fc_{t_{N+1-i}})$ for $1 \le i \le k+1$.
Note that, according to Assumption \ref{ass:R} (iii), the particular case $i=k+1$ gives 
\begin{equation}\label{eq:lim1}
\lim_{n\uparrow \infty}R_{t_{N-k}}(\we \beta^n_{t_{N-(k+1)}}) - \we \beta^n_{t_{N-(k+1)}}= R_{t_{N-k}}(\beta^0_{t_{N-(k+1)}}) - \beta^0_{t_{N-(k+1)}} \; .
\end{equation}
The general case $i\le k$ follows from Assumption \ref{ass:R} (ii) applied to equation \reff{eq:limite de R}:
the left hand term converges to $0$ and $(1-\frac{2 \| C\|}{1+\eta^n})$ converges to $1$, so that
\begin{equation}\label{eq:lim2}
\lim_{n\uparrow \infty}R_{t_{N+1-i}}(\we \beta^n_{t_{N-i}}) - \we \beta^n_{t_{N-i}}+ \ell_{t_{N+1-i}}^n = R_{t_{N+1-i}} (0)  \; .
\end{equation}
By construction of the subsequence, 
the convexity of $\Xf_{t_{N-(k+1)}, adm}^0 (T)$ and the belonging of $0$ to that set,
$\we \xi^n_T \in \Xf_{t_{N-(k+1)}, adm}^0 (T)$.
By using property of Assumption \ref{ass: portefeuille} (ii) and 
since the sequence $(\ell^n_{t_{N+1-i}})_{n\ge 1} $ is uniformly bounded for any $1\le i \le k+1$, 
see proof of Proposition \ref{prop:FC1} above, we define
$$
\wt V^n_T := \we \xi^n_T
 +\ell^n_{t_{N-k}} + \sum_{i=1}^k\left(  R_{t_{N+1-i}} (\we \beta^n_{t_{N-i}}) - 
 \we \beta^n_{t_{N-i}}+\ell^n_{t_{N+1-i}}\right)  \in \we \Xf^k_{t_{N-(k+1)}} (T) \; ,
$$
which converges by definition and equations 
\reff{eq:lim1} and \reff{eq:lim2} to $\wt V^0_T$ such that
\begin{equation}\label{eq:better than}
\wt V^0_T +R_{t_{N-k}} (\beta^0_{t_{N-(k+1)}}) - \beta^0_{t_{N-(k+1)}}  
\succeq_T\sum_{i=1}^{k+1} R_{t_{N+1-i}} (0) \; .
\end{equation}
Notice also that by Assumption \ref{ass:R} (ii), for all $n\ge1$
$$
\wt V_T^n = \we V_T^n -R_{t_{N-k}}(\we \beta^n_{t_{N-(k+1)}}) + \we \beta^n_{t_{N-(k+1)}} +  \sum_{i=1}^{k+1}\left(1- \frac{2\|C\|}{1+\eta^n}\right)R_{{t_{N+1-i}}} (0) \succeq_T -(\kappa + (k+1)\Kf) \; .
$$
By Fatou-closedness of $\we \Xf^k_{t_{N-(k+1)}}(T)$,
we finally obtain that 
$ \wt V^0_T + R_{t_{N-k}}(\beta^0_{t_{N-(k+1)}}) - \beta^0_{t_{N-(k+1)}} \in \we \Xf^{k+1}_{t_{N-(k+1)}}(T)
$.
By equation \reff{eq:better than} and {\bf CSP}(R), 
$\|\beta^0_{t_{N-(k+1)}}\| \le C$ but by construction, 
$\|\beta^0_{t_{N-(k+1)}} \|=2\|C\|$, so that we fall on a contradiction.
The case {\bf 2.} is not possible.
\ep

\vs2

Remark that the flexibility of the {\bf CSP}(R) condition is reflected in 
the construction in equation \reff{eq:normalization} used in the last lines
of the proof of Proposition  \ref{prop:FC2}.
The choice of a good norm for $\we \beta$ can indeed vary according to 
the condition we aim at. 
Following Propositions \ref{prop:FC1} and \ref{prop:FC2}, 
$\we \Xf^{k+1}_{t_{N-(k+1)}}(T)$ is Fatou-closed if $\we \Xf^{k}_{t_{N-k}}(T)$ is Fatou-closed.
Proposition \ref{prop:FC1} is used a last time to pass from 
the closedness of $\we \Xf^N_{t_0}(T)$ to the closedness of $\Xf^R_{0}(T)$.
%
%
%

\subsection{Proof of Theorem \ref{th:SR}}

\proof
The ``$\Rightarrow$'' sense is obvious. 
To prove the ``$\Leftarrow$'' sense, 
we take $H\in L^0(\R^d, \Fc_T)$ such that $H\succeq -\kappa$
for some $\kappa \in \R^d_+$ and such that $\Esp{Z H}\le \alpha^R_0(Z)$ 
for all $Z\in \Mc$ and $H\notin \Xf_{0}^R(T)$, 
and work toward a contradiction.
Let $(H^n)_{n\ge1}$ be the sequence defined by 
$
H^n := H \ind{\|H\| \le n} - \kappa \ind{\| H\| > n}
$. 
By Proposition \ref{prop:FC}, $\Xf^R_{0}(T)$ is Fatou-closed, 
so by Lemma 5.5.2 in \cite{kabanov2010book}, 
$\Xf^R_{0} (T)\cap L^\infty(\R^d, \Fc_T)$ is weak*-closed.
Since $H\notin \Xf_{0}^R(T)$, 
there exists $k$ large enough such that 
$
H^k \notin \Xf_{0}^R(T)\cap L^\infty(\R^d, \Fc_T)$ 
but, 
because any $Z\in \Mc$ has positive components,
still satisfies
\begin{equation}\label{eq:contradic}
\Esp{Z'_T H^k}
\le \alpha^R_{0}(Z)
:=\sup\brace{\Esp{Z'_T V_T}~:~V_T \in \Xf^R_{0}(T)}
\quad \pourtout Z\in \Mc.
\end{equation}
By Proposition \ref{prop:convexity}, 
the set $\Xf_{0}^R(T)$ is convex, 
so that we deduce from the Hahn-Banach theorem that 
we can find $z\in L^1(\R^d, \Fc_T)$  such that
\begin{equation}\label{eq:contradiction}
\sup \left\{\Esp{z' V_T} ~:~ V_T \in \Xf_{0}^R(T)\cap L^\infty(\R^d, \Fc_T)\right\} < \Esp{z' H^k}<+\infty.
\end{equation}
We define $\we Z$ by $\we Z_t = \Esp{z| \Fc_t}$.
By using the same argument as in Lemma 3.6.22 in \cite{kabanov2010book}, 
we have that 
$\Xf_{0}^R (T) \cap L^\infty (\R^d, \Fc_T)$ is dense 
in $\Xf_{0}^R (T)$ and
so that the left hand term of 
equation \reff{eq:contradiction} is precisely 
$\alpha_0^R(\we Z)$. 
The process $\we Z$ is a non negative martingale and since 
$$
\left (\Xf_{0}^R (T)  - L^\infty (\wt K_{t}, \Fc_t)\right )\subset \left (\Xf_{0}^R (T) \cap L^\infty (\R^d, \Fc_T)\right ) \ \forall t\in [0,T] \; , 
$$
we have $\we Z_t \in L^1 (\wt K^*_t, \Fc_t)$.
The contrary would make the left term of equation \reff{eq:contradiction} 
equal to $+\infty$ for
suitable sequences $(\xi^m)_{m\ge 1}\subset \Xf^R_{0}$
(see the proof of Proposition 3.4 in \cite{bouchard2011nma}).
By using the same arguments as above, 
and since $\Xf^0_{0,adm}(T)$ is Fatou-closed too,
we have that $\Xf^0_{0,adm}(T)\cap L^\infty (\R^d, \Fc_T)$ is dense in $\Xf^0_{0,adm}(T)$ .
This implies that
\begin{align*}
\alpha^0_0(\we Z) &:= \sup\brace{\Esp{\we Z'_T V_T} ~:~ V_T \in \Xf_{0, adm}^0(T)}\\
& =  \sup\brace{\Esp{\we Z'_T V_T} ~:~ V_T \in \Xf_{0, adm}^0(T)\cap L^\infty(R^d, \Fc_T)}\\
& \ge \sup\brace{\Esp{\we Z'_T V_T} ~:~ V \in \Xf_{0}^0 \And V_\tau \succeq_\tau -\kappa  \pourtout \tau\in \Tc, \text{ for some }\kappa\in \R^d_+}
\end{align*}
Moreover, according to Assumption \ref{ass:R} (ii),
$\xi_T + \sum_{i=1}^N R_\ti (0) \in \Xf_{0}^R(T)\cap L^\infty(\R^d, \Fc_T)$ for any $\xi_T \in \Xf_{0, ad�}^0(T)\cap L^\infty(\R^d, \Fc_T)$, 
so that
$$
\alpha^0_{0}(\we Z)  - N \we Z'_0 \Kf
\le \alpha^0_{0}(\we Z)  + \Esp{\we Z'_T \sum_{i=1}^N R_\ti (0)}
\le \sup \left\{\Esp{z' V_T} ~:~ V_T \in \Xf_{0}^R(T)\cap L^\infty(\R^d, \Fc_T)\right\} 
$$
and then $\alpha^0_{0}(\we Z)$ is finite according to equation \reff{eq:contradiction}.
Take $Z\in \Mc$.
Then there exists $\eps>0$ small enough such that, 
by taking $\check{Z} = \eps Z + (1-\eps)\we Z$,
$$
\alpha^R_0(\check{Z}) \le \eps\alpha^R_0(Z)+ (1-\eps)\alpha^R_0(\we Z)< \eps\Esp{Z'_T H^k}+(1-\eps)\Esp{\we Z'_T H^k} = \Esp{\check{Z}'_T H^k}\; .
$$
It is easy to see that $\check{Z} \in \Mc$, 
so that the above inequality contradicts \reff{eq:contradic}.
\ep

\vs2

{\bf Aknowledgement}: the author wants to thank the Editor and anonymous referee for
their presentation advice. He also thanks Bruno Bouchard for providing leading ideas and careful reading which greatly improve this paper.

\end{document}